\documentclass[11pt,a4paper]{article}
\usepackage[pdftex]{graphicx}

\usepackage{amssymb}

\usepackage{epsfig}
\usepackage{amsmath}
\usepackage{amssymb}
\usepackage{natbib}

\newcommand\pd[3][]{\frac{\partial^{#1} #2}{\partial{#3}^{#1}}}
\newcommand{\bs}{\boldsymbol}
\newcommand{\smum}{\, \mu \mathrm{m}}

\begin{document}
\title{\textsf{\textbf{A boundary element regularised Stokeslet method applied to cilia and flagella-driven flow}}\footnote{Postprint of an article published in: Proceedings of the Royal Society of London Series A, Vol. 465, 3605--3626, (2009). The published article can be found at http://rspa.royalsocietypublishing.org/content/465/2112/3605}}
\author{\textbf{David J. Smith}\footnote{Email address: d.j.smith.2@bham.ac.uk} \\
\small{School of Mathematics and School of Clinical and Experimental Medicine,}\\
\small{University of Birmingham, Edgbaston, Birmingham B15 2TT, United Kingdom} \\
\small{Centre for Human Reproductive Science, Birmingham Women's NHS Foundation Trust,}\\ 
\small{Metchley Park Road, Edgbaston, Birmingham B15 2TG, United Kingdom}
} 
\date{}


\maketitle

\begin{abstract}
A boundary element implementation of the regularised Stokeslet method of Cortez is applied to cilia and flagella-driven flows in biology. 
Previously-published approaches implicitly combine the force discretisation and the numerical quadrature used to evaluate boundary integrals.
By contrast, a boundary element method can be implemented by discretising the force using basis functions, and calculating integrals using accurate
numerical or analytic integration. This substantially weakens the coupling of the mesh size for the force and the regularisation parameter, and greatly reduces
the number of degrees of freedom required.
When modelling a cilium or flagellum as a one-dimensional filament, the regularisation parameter can be considered
a proxy for the body radius, as opposed to being a parameter used to minimise numerical errors.
Modelling a patch of cilia, it is found that: (1) For a fixed number of cilia, reducing cilia spacing reduces transport. (2) For fixed patch dimension, increasing cilia number increases the transport, up to a plateau at $9\times 9$ cilia. Modelling a choanoflagellate cell it is found that the presence of a lorica structure significantly affects transport and flow outside the lorica, but does not significantly alter the force experienced by the flagellum.
\end{abstract}

\section{Introduction}
Fluid dynamic phenomena on microscopic scales are fundamental to life, from the feeding and swimming of bacteria and single-celled eukaryotes to complex roles in the organ systems of mammals. Examples include sperm swimming, ovum and embryo transport, respiratory defence, auditory perception and embryonic symmetry-breaking. Cilia, which range from a few to tens of microns in length, are hair-like organelles project from cell surfaces either in dense mats or one-per-cell, and through specialised side-to-side or rotational beating patterns and viscous interaction with the cell surface, propel fluid in a parallel direction. Flagella may be tens to hundreds of microns in length,
typically occurring singly or two per cell, and generally produce propulsion tangential to the flagellum axis through the propagation of a bending wave, resulting in swimming as in sperm, or the generation of feeding currents as in \textit{choanoflagellates}. These phenomena have since the work of Taylor, and Gray and Hancock in the 1950s, motivated the study of methods for solving the zero Reynolds number linear 
`Stokes flow' equations,
\begin{eqnarray}
0 & = & - \bs{\nabla} p + \mu \nabla^2 \bs{u} + \bs{f}   \nonumber \\
\mbox{and} \quad \quad \bs{\nabla} \cdot \bs{u} & = & 0 \mbox{,} \label{stokes}
\end{eqnarray}
where $\bs{u}$, $p$, $\mu$ and $\bs{f}$ are velocity, pressure, kinematic viscosity and force per unit volume respectively. These equations model flow strongly dominated by viscous effects with negligible inertia, valid for cilia and flagella since the Reynolds number can be estimated as $\mathrm{Re}=\mu UL/\rho<10^{-2}$. \cite{hancock53}, working with Sir James Lighthill, introduced the singular `Stokeslet' solution, corresponding to a concentrated point-force, or equivalently the purely viscous component of the flow driven by a translating sphere. The Stokeslet is the solution of the Stokes flow equations 
for unit force acting in the $j$-direction and concentrated at $\bs{\xi}$, corresponding to taking $\bs{f}(\bs{x})=\delta(\bs{x}-\bs{\xi}) \bs{e}_j$, where $\delta(\bs{x}-\bs{\xi})$ denotes the Dirac delta distribution and $\bs{e}_j$ is the appropriate basis vector. The $i$-component of the velocity field driven by this force is written as $S_{ij}(\bs{x},\bs{\xi})$, and in an infinite fluid takes the form
\begin{equation}
S_{ij}(\bs{x},\bs{\xi})=\left(\frac{\delta_{ij}}{r} + \frac{r_i r_j}{r^3}\right) \mbox{,}
\end{equation}
where $r_i=x_i-\xi_i$ and $r^2=|\bs{x}-\bs{\xi}|^2=r_1^2+r_2^2+r_3^2$, with $\delta_{ij}$ denoting Kronecker delta tensor. The flow due to a force $\bs{F}$ concentrated at the point $\bs{\xi}$ 
corresponds to taking $\bs{f}(\bs{x})=\delta(\bs{x}-\bs{\xi})\bs{F}$ in equation~\eqref{stokes}, the solution being given by
$u_i(\bs{x})=(1/8\pi\mu)S_{ij}(\bs{x},\bs{\xi})F_j$. We assume the summation convention over repeated indices throughout.

The Stokeslet forms the basis for slender body theory for Stokes flow \citep{burgers38,hancock53,tuck64,batchelor70,johnson}, its local simplification `resistive force theory', introduced by 
\citet{grayhan} 
and the single-layer boundary integral method for Stokes flow \citep[see for example][]{pozrikidis,bemlib}. These techniques have proved very useful in the modelling of cilia-driven flows 
\citep{bl72,liron76,gueron92,smithdsl}, sperm motility \citep{dresdner,smithjfm}, bacterial swimming \citep{higdon79he,phan,ramia} and atomic force microscopy \citep{clarke}, allowing three-dimensional flow problems with moving boundaries to be solved with greatly reduced computational cost compared with direct numerical solutions of the differential equations~\eqref{stokes}. These techniques involve expressing the
fluid velocity field with an integral equation of the form
\begin{equation}
\bs{u}(\bs{x})=\frac{1}{8\pi\mu} \int_S \bs{f}(\bs{\xi}) \cdot \bs{\mathsf{S}}(\bs{x},\bs{\xi}) \; \mathrm{d}S_{\bs{\xi}} \mbox{,} \label{intEq}
\end{equation}
where $S$ is a collection of surfaces or lines, for example representing boundaries, cell surfaces, cilia or flagella, with the symbol $\int_S$ denoting
line or surface integrals as appropriate, and $\bs{f}(\bs{\xi})$ denoting force per unit area or length respectively. The force exerted by the body on the fluid is given by $\bs{f}(\bs{\xi}) \mathrm{d}S_{\bs{\xi}}$, and the force exerted by the fluid on the body is given by$-\bs{f}(\bs{\xi})\mathrm{d}S_{\bs{\xi}}$.
The force density is calculated by combining equation~\eqref{intEq} 
with either prescribed motions of the bodies, or a coupled elastic/active force model, through the no-slip, no-penetration condition $\bs{u}(\bs{\xi})=\dot{\bs{\xi}}$.

\section{The method of regularised Stokeslets}
Line distributions of Stokeslets, for example equation~\eqref{intEq} with the boundary $S$ parametrised as $\bs{\xi}(s)$ for scaled arclength parameter $0<s<1$, representing objects such as cilia or flagella, have the principal disadvantage that the flow field is singular at any point $\bs{x}=\bs{\xi}(s)$.
To calculate the force per unit length, it is necessary to perform collocation at points on the `surface' of the slender body, displaced a small distance from the centreline $\bs{X}(s_q)=\bs{\xi}(s_q)+a(s_q)\bs{n}(s_q)$,
where $a(s_q)$ is equivalent to the slender body radius and $\bs{n}(s_q)$ is a unit normal vector. The velocity field is regular in the region designated the outside of the body, with the singular line distribution lying inside the notional surface of the cilium.

Point distributions of Stokeslets $\sum_{q=1}^N \bs{F}_q \cdot \bs{\mathsf{S}}(\bs{x},\bs{\xi}_q)$, for example representing a swarm of cells, are also singular at any $\bs{x}=\bs{\xi}_q$. Surface distributions of Stokeslets, as used in the single-layer boundary integral method for Stokes flow, do not result in singular velocities, but nevertheless require careful numerical implementation in order to ensure that the surface integrals are calculated correctly \citep{pozrikidis,bemlib}.

In order to circumvent these difficulties and ensure a regular flow field throughout the flow domain, \citet{cortez01} 
introduced the `regularised Stokeslet'. This is defined as the exact solution to the Stokes flow equations with smoothed point-forces,
\begin{eqnarray}
0 & = & -\bs{\nabla}p+\mu\nabla^2\bs{u}+\bs{f}\psi_\epsilon(\bs{x}-\bs{\xi}) \mbox{,} \nonumber \\
\mbox{and} \quad \quad \bs{\nabla}\cdot\bs{u}  & = & 0        \mbox{.}
\end{eqnarray}
The symbol $\psi_\epsilon$ denotes a cutoff-function or `blob' with regularisation parameter $\epsilon$, satisfying $\int_{\mathbb{R}^3} \psi_\epsilon(\bs{x})\mathrm{d}V_{\bs{x}}=1$. \cite{cortez05} showed that 
with the choice $\psi_\epsilon(\bs{x}-\bs{\xi}):=15\epsilon^4/(8\pi\mu r_\epsilon^7)$, the regularised Stokeslet velocity tensor is given by
\begin{equation}
S_{ij}^\epsilon(\bs{x},\bs{\xi})=\frac{\delta_{ij}(r^2+2\epsilon^2)+r_i r_j}{r_\epsilon^3} \mbox{.} \label{regSto}
\end{equation}
Here and in the rest of the paper, we use the compact notation $r_\epsilon=\sqrt{r^2+\epsilon^2}$. 
We follow Cortez and co-authors in using the solution $S_{ij}^\epsilon$, and its counterpart near a no-slip boundary $B_{ij}^\epsilon$ given in equation~\eqref{ainleyimage}, as the basis for our computational study.

\citet{cortez05} derived the equivalent Lorentz reciprocal relation, and hence a boundary integral equation for the RSM. This leads to the following equation for the fluid velocity at location $\bs{x}$, where the surfaces and lines representing the cells and beating appendages are denoted $S$,
\begin{equation}
\bs{u}(\bs{x})=\frac{1}{8\pi\mu}\int_S \bs{f}(\bs{\xi}) \cdot \bs{\mathsf{S}}^\epsilon(\bs{x},\bs{\xi}) \; \mathrm{d}S_{\bs{\xi}} \mbox{.} \label{rsIntVel}
\end{equation}
As above, the unknown $\bs{f}(\bs{\xi})$ is a force per unit area or length depending on whether $\bs{\xi}$ is on a surface or line comprising $S$.
A significant advantage of the RSM
is that the kernel remains regular for $\bs{x}=\bs{\xi}$, even when
$S$ consists of lines or points rather than surfaces,
and so collocation can be performed at such points without the need to make a finite displacement.
This method has proved effective in modelling a number of biological flow problems, including flow due to an individual cilium and the swarming of bacteria \citep{ainley08}, the bundling of bacterial flagella \citep{flores05,cisneros08}, hydrodynamic interaction of swimming cells \citep{cisneros07} and the flagellar motility of human sperm \citep{gillies09}.

\section{Alternative implementation of the RSM as a boundary element method}
The mathematical problem considered in this paper is the calculation of the force density $\bs{f}(\bs{\xi})$ from known boundary velocity, given by the time derivative $\dot{\bs{\xi}}$, through applying the no-slip condition $\bs{u}(\bs{\xi})=\dot{\bs{\xi}}$ to equation~\eqref{rsIntVel}.
The boundary, representing a cell surface, a flagellum, an array of cilia, a network of fibres or a collection of particles, is 
discretised as a set of points $\{\bs{\xi}_q:q=1,\ldots,N\}$. Applying collocation at these points gives the following equation
\begin{equation}
\bs{u}(\bs{\xi}_q)=\frac{1}{8\pi\mu}\int_S \bs{f}(\bs{\xi}) \cdot \bs{\mathsf{S}}^\epsilon(\bs{\xi}_q,\bs{\xi}) \; \mathrm{d}S_{\bs{\xi}} \mbox{.}
\end{equation}

The implementation of the regularised Stokeslet method used in \cite{cortez05,ainley08,gillies09,cisneros07} is equivalent to approximating the force density by constant values, $\bs{f}(\bs{\xi}_r) \approx \bs{f}_r$, and replacing the integral of the Stokeslet with a low order quadrature, giving
\begin{equation}
\bs{u}(\bs{\xi}_q)=\frac{1}{8\pi\mu}\sum_{r=1}^N w_r \bs{f}_r \cdot \bs{\mathsf{S}}^\epsilon(\bs{\xi}_q,\bs{\xi}_r)  \mbox{,} \label{simpleImp}
\end{equation}
where $\{w_1,\ldots,w_N\}$ are quadrature weights. Hence the matrix equation $AX=B$ can be formed, where
\begin{eqnarray}
A_{3(q-1)+i,3(r-1)+j}&=& w_r S_{ij}^\epsilon(\bs{\xi}_q,\bs{\xi}_r)/(8\pi\mu)  \mbox{,} \nonumber \\
X_{3(r-1)+j}&=& (\bs{f}_r)_j \nonumber \\
\mbox{and} \quad \quad B_{3(q-1)+i} & = & u_i(\bs{\xi}_q). \label{matrixSetupSimple}
\end{eqnarray}

This discretisation can be viewed as a constant-force implementation of a boundary integral method, utilising a low order quadrature in which the abscissae are identified with
the collocation points. 
However, while the force density $\bs{f}(\bs{\xi})$ varies relatively slowly, the kernel $\bs{\mathsf{S}}^\epsilon(\bs{\xi}_q,\bs{\xi})$ will vary rapidly in the neighbourhood of $\bs{\xi}_q$.
Hence if the surface is discretised as patches $S_1,\ldots,S_N$,
the term $w_r\bs{f}_r \cdot \bs{\mathsf{S}}^\epsilon(\bs{\xi}_q,\bs{\xi}_r)$ in equation~\eqref{simpleImp} may not be an accurate approximation to
$\int_{S_r} \bs{f}(\bs{\xi}) \cdot \bs{\mathsf{S}}^\epsilon(\bs{\xi}_q,\bs{\xi})\; \mathrm{d}S_{\bs{\xi}}$, particularly when $r=q$. 

Consequently, in order to obtain accurate solutions, it is necessary to use a relatively large number of nodes $\bs{\xi}_q$, many degrees of freedom, and hence a large dense matrix system in order to calculate accurate solutions, as
discussed in detail in \ref{testing}(\ref{sphere}). This is an unnecessary computational effort, since even with a constant-force discretisation, far fewer degrees of freedom for $\bs{f}$ are required than quadrature nodes for the evaluation of the kernel integrals.
It is also 
necessary to employ an empirical rule of the form $\epsilon=C(\delta s)^m$ relating the regularisation parameter to the mesh size $\delta s$ to obtain the expected result, as discussed in \citet{cortez05,ainley08}.

For this reason, we instead implement a constant-force boundary element method where the integration and force discretisation are `decoupled', which we then apply to a number of example problems. On each patch $S_r$, the force is approximated by 
$\bs{f}_r$. Then we have
\begin{equation}
\bs{u}(\bs{\xi}_q)=\frac{1}{8\pi\mu}\sum_{r=1}^N \bs{f}_r\cdot \int_{S_r}  \bs{\mathsf{S}}^\epsilon(\bs{\xi}_q,\bs{\xi}) \; \mathrm{d}S_{\bs{\xi}}. \label{refinedImp}
\end{equation}
This problem can again be written as a matrix equation $AX=B$, with $A$ now taking the form
\begin{equation}
A_{3(q-1)+i,3(r-1)+j}=\frac{1}{8\pi\mu}\int_{S_r} S_{ij}^\epsilon(\bs{\xi}_q,\bs{\xi}) \; \mathrm{d}S_{\bs{\xi}}. \label{matrixSetup}
\end{equation}
The $\epsilon$--$\delta s$ coupling of the original method is now substantially weakened, for the following reason: the discretisation chosen for the force no longer has any bearing on the evaluation of the kernel integrals in equation~\eqref{matrixSetup}. If numerical integration is used it is still necessary to choose a sufficiently refined method for the value of $\epsilon$ chosen, for the `near-singular' integrals. However, for the values of $\epsilon$ typically used, this is far less computationally costly than increasing the number of degrees of freedom, as shown in \ref{testing}, and moreoever may be accomplished in principle with an automatic integration routine or in some cases analytic integration.

The choice of constant-force elements is purely for simplicity: higher-order methods may also be used, generalising the above so that $\bs{f}(s)$ is approximated as $\sum_{r=1}^N \phi_r(s)\bs{f}_r$, where $\{\phi_1(s),\ldots,\phi_N(s)\}$ is a set of basis functions. Cubic Lagrange interpolating polynomials were used by \citet{smithvecil} in the context of the non-regularised viscoelastic Stokeslet distributions.
The integrals $\int_{S_r}  \bs{\mathsf{S}}^\epsilon(\bs{\xi}_q,\bs{\xi}) \; \mathrm{d}S_{\bs{\xi}}$ may be evaluated using any appropriate means: Gauss-Legendre numerical quadrature, more advanced adaptive quadratures, or in the case of problems where the 
boundary is reduced to a line segment, the integrals may be performed analytically (see \ref{appendixAnalytic}). In \ref{testing} we compare the two approaches \eqref{simpleImp} and \eqref{refinedImp}.

\section{The Ainley et al. image system for a no-slip plane boundary}
In order to model fluid propulsion by cilia protruding from a cell surface, \cite{bl71sto} derived the Stokeslet image system satisfying 
the no-slip, no-penetration condition $\bs{u}=0$ on $x_3=0$.
The technique is equally useful in modelling the interaction of a glass surface and a sperm cell, and removes the need for the cell surface boundary to be included in the boundary integral equation. We shall not repeat the solution here, however it consists of a Stokeslet in the fluid, an equal and opposite image Stokeslet, a Stokes-dipole and potential dipole. 
As an important step in generalising the regularised Stokeslet method to the modelling of biological flows, \cite{ainley08} derived the equivalent regularised image 
system, which we denote $B_{ij}^\epsilon$ for flow near a no-slip boundary. We rewrite the solution in index notation for direct comparison with the result of \cite{bl71sto}:
\begin{eqnarray}
B_{ij}^\epsilon(\bs{x},\bs{\xi}) & = & \frac{\delta_{ij}(r^2+2\epsilon^2)+r_i r_j}{r_\epsilon^3}   		                  
       - \frac{\delta_{ij}(R^2+2\epsilon^2)+R_i R_j}{R_\epsilon^3}               \nonumber \\
& &            + 2 h  \Delta_{jk} \left[ \pd{}{R_k} \left( \frac{h R_i}{R_\epsilon^3} -   \frac{\delta_{i3}(R^2+2\epsilon^2)+R_i R_3}{R_\epsilon^3} \right) -4\pi h \delta_{ik} \phi_\epsilon(R) \right] 
\nonumber \\
& & -\frac{6h\epsilon^2}{R_\epsilon^5}(\delta_{i3}R_j-\delta_{ij}R_3) 
\mbox{.} \label{ainleyimage}
\end{eqnarray}
The tensor $\Delta_{jk}$ takes value $+1$ for $j=k=1,2$, value $-1$ for $j=k=3$ and zero otherwise, originally 
written as $(\delta_{j\alpha} \delta_{\alpha k} - \delta_{j3} \delta_{3k})$ by Blake.
The last line of equation~\eqref{ainleyimage} is not precisely equivalent to that in \cite{ainley08}, the latter having a typographical sign error. The term $\phi_\epsilon(R):=3\epsilon^2/(4\pi R_\epsilon^5)$ is a more slowly-decaying
blob than $\psi_\epsilon(R)$, discovered by \cite{ainley08} as the function generating the potential dipole appropriate to the no-slip image system. 
The additional term on the final line is the difference between rotlets arising from the two different blobs $\psi_\epsilon(R)$ and $\phi_\epsilon(R)$, and is 
$O(\epsilon^2/R^4)$.


\begin{figure}
$
\begin{array}{ll}
  \mbox{(a) cilium }(2,2)\mbox{, timestep }37/40
  & 
  \mbox{(b) cilium }(3,1)\mbox{, timestep }11/40
  \\
  {\includegraphics{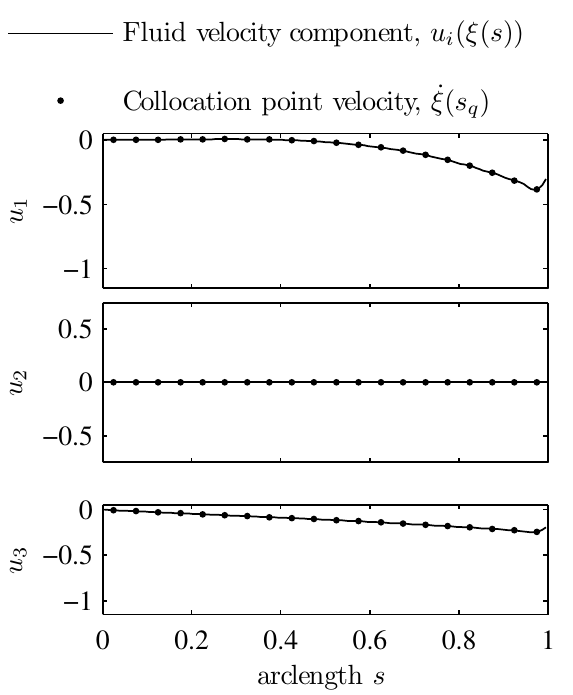}}
  &
  {\includegraphics{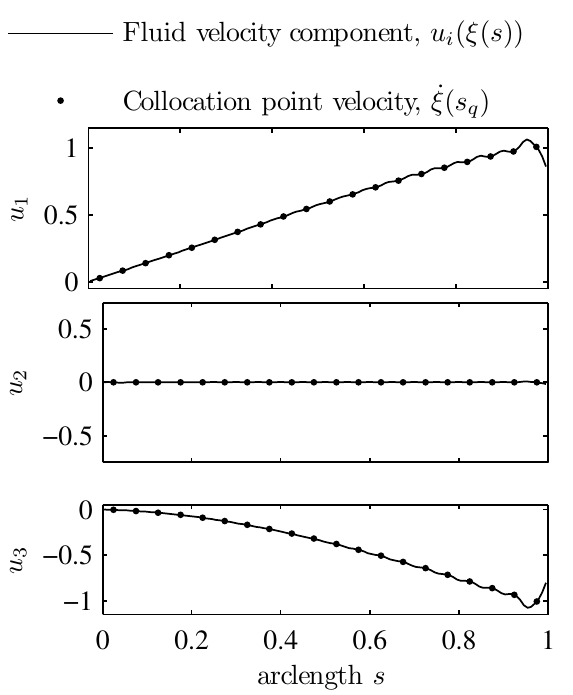}}
\end{array}
$
\caption{Post-hoc verification of the accuracy of the solution, comparing the fluid velocity on the cilium $\bs{u}(\bs{\xi}(s,t_n))$ for dimensionless arclength $0<s<1$ and the prescribed cilium velocity on the collocation points $\dot{\bs{\xi}}(s_q,t_n)$ for $q=1,\ldots,N$, at two different timesteps $t_n$. Results were computed with a square lattice of $3\times 3$ cilia with spacing $0.08L$, beating in synchrony, for one beat cycle, with $N=20$ collocation points and $N_T=40$ timesteps per cycle. Results are shown for (a) an `average' case, cilium $(2,2)$, timestep $37/40$, for which the absolute RMS error is $0.0132$, and (b) the worst case in the beat cycle, on cilium $(3,1)$, timestep $11/40$, for which the error is $0.0357$. All velocities are given in dimensionless units, scaled with respect to $\sigma L$.
}\label{vCilTest}
\end{figure}

\begin{figure}
\[
\begin{array}{ll}
  \mbox{(a), timestep }10/40 
  & 
  \mbox{(b), timestep }20/40
  \\
  \begin{array}{c}
  \epsfig{file=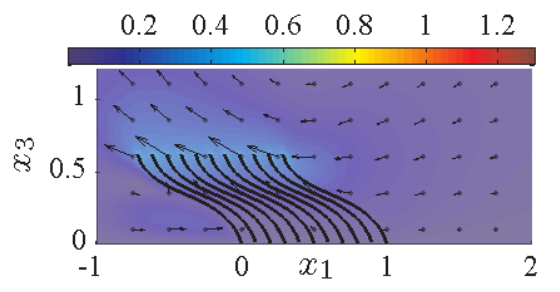} \\ 
  \epsfig{file=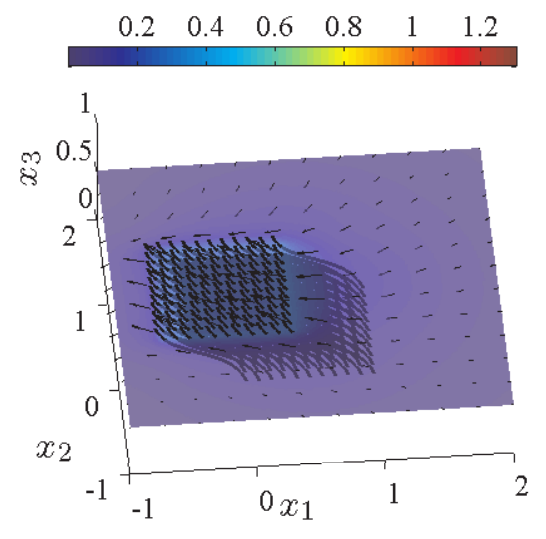} 
  \end{array}
  &
  \begin{array}{c}
  \epsfig{file=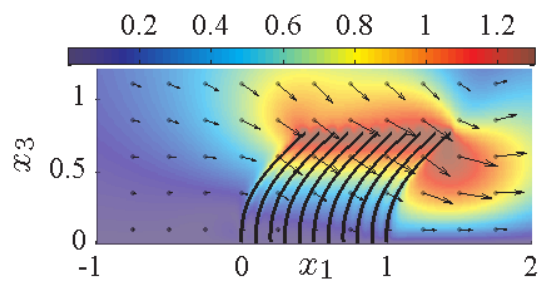} \\ 
  \epsfig{file=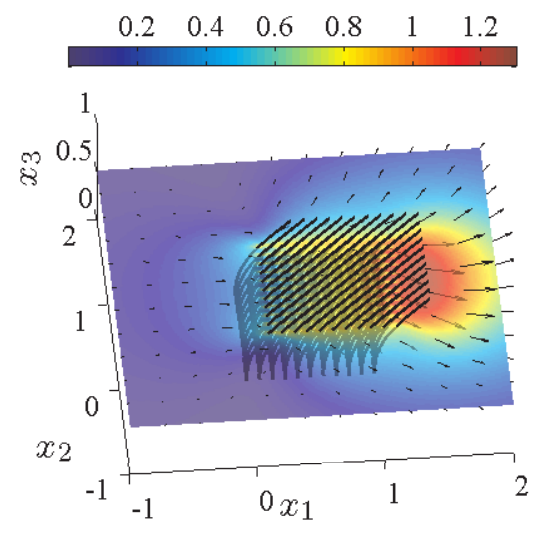} 
  \end{array}
  \\
  \mbox{(c), timestep }30/40
  & 
  \mbox{(d), timestep }40/40 
  \\
  \begin{array}{c}
  \epsfig{file=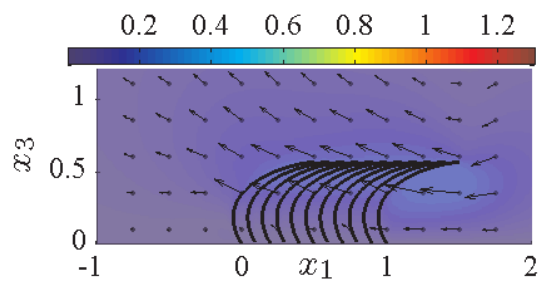} \\ 
  \epsfig{file=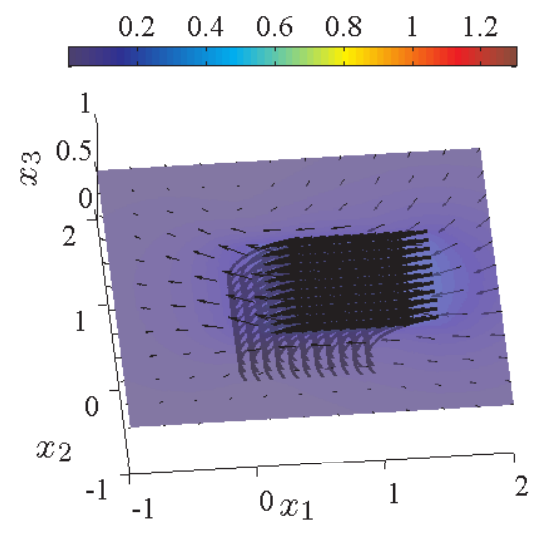} 
  \end{array}
  &
  \begin{array}{c}
  \epsfig{file=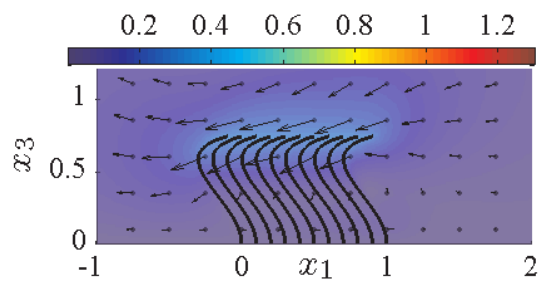} \\ 
  \epsfig{file=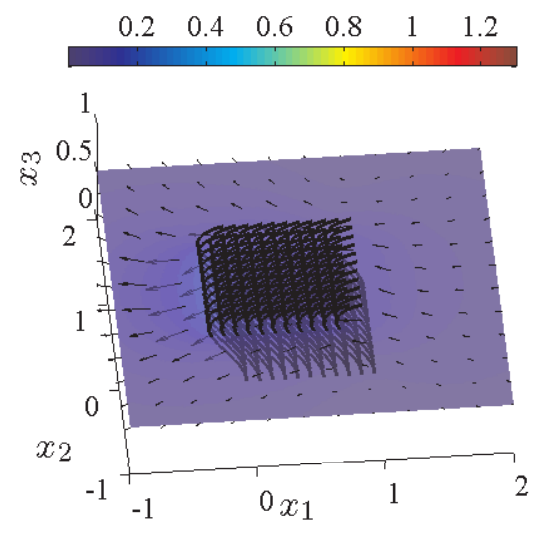} 
  \end{array}
\end{array}
\]
\caption{Velocity field results for a square patch of $121$ cilia, with (a-d) showing timesteps $10/40$, $20/40$, $30/40$ and $40/40$ over one complete beat cycle. The upper panels show a `vertical' section of the velocity field at $x_2=0.5$; the lower panels show a `horizontal' section at $x_3=0.5$. Velocity magnitude is shown in colour, direction is shown with arrows.
}\label{vFieldCilia}
\end{figure}

\begin{figure}
$
\begin{array}{c}
  \mbox{(a), fixed cilia number }3\times 3\mbox{, variable cilia spacing}       \\
  {\includegraphics[angle=90]{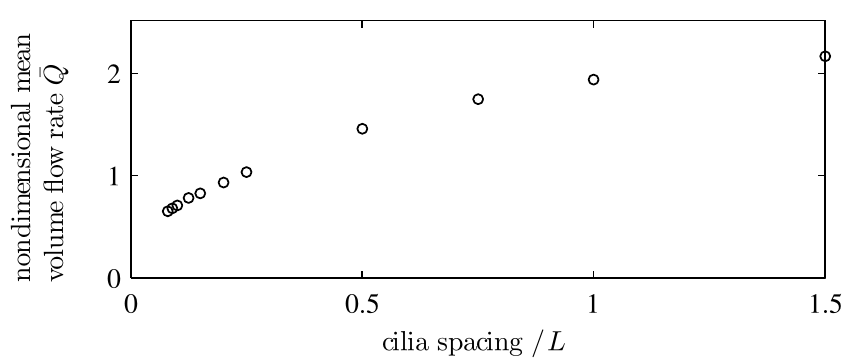}}        \\
  \mbox{(b), fixed patch dimension }1.0L\times 1.0L\mbox{, variable cilia number} \\
  {\includegraphics[angle=90]{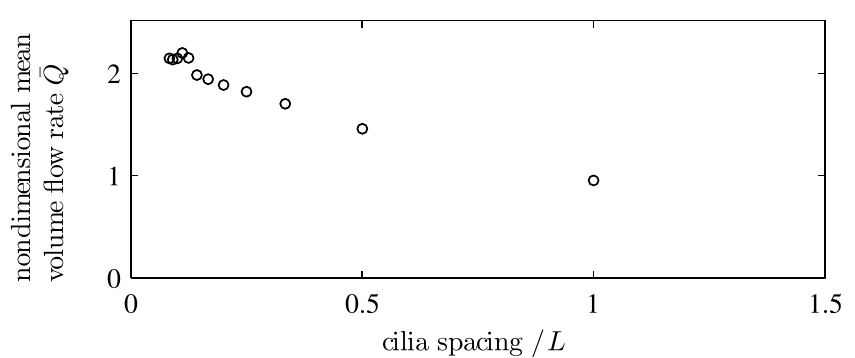}}
\end{array}
$
\caption{Mean volume flow rate $\bar{Q}$ versus cilia spacing, (a) for a patch of $3\times 3$ cilia, the patch size varying from $0.16L\times 0.16L$ to $3.0L\times 3.0L$,
(b) for a patch of fixed size $1.0L\times 1.0L$, the cilia numbers varying from $2\times 2$ to $13\times 13$.}\label{vfrFig}
\end{figure}

\section{Application to biological flows}\label{biological}
The numerical schemes given by equations~\eqref{matrixSetupSimple} and \eqref{matrixSetup} are applied to test problems involving rod and sphere translation in an infinite fluid in \ref{testing}. In this section we apply the method to two biologically-inspired problems: determining the flow generated by an array of cilia protruding from a flat surface, and by a choanoflagellate cell in an infinite fluid surrounded by a silica lorica structure.
We nondimensionalise with scalings $L$, $1/\sigma$ and $\sigma L$ for length, time and velocity, where $L$ is cilium or flagellum length and $\sigma$ is radian beat frequency.
The appropriate scaling for force per unit length on a filament is $\mu \sigma L$, and for force per unit area on a surface is $\mu \sigma$.
\subsection{Modelling a patch of cilia}\label{cilia}
Cilia lining the airway epithelia and the surfaces of microorganisms typically appear in dense `fields', while cilia within the female reproductive tract generally appear in patches, with ciliated cells being interspersed between secretory cells. The density of cilia in such systems can be very high, approximately $6$--$8 / \mu\mathrm{m}^2$ \citep{sleigh88}, which for a square lattice is equivalent to a spacing of $0.35$--$0.41 \smum$, or approximately $0.06L$--$0.07L$ for cilium length $L=6\smum$. This contrasts with embryonic nodal primary cilia, for which there is only one cilium per cell. 

Slender body theory does not in general produce exact solutions, except in certain special cases such as that of a straight ellipsoids in a uniform flow, or uniform shear \citep{chwang}. When modelling single bodies with non-uniform velocity or curvature, or multiple bodies, the flow field $\bs{u}(\bs{x},t_n)$ generated by a numerical solution for $\bs{f}(s,t_n)$ typically exhibits errors when
comparing $\bs{u}(\bs{\xi}(s,t_n))$ with the prescribed boundary movement $\dot{\bs{\xi}}(s,t_n)$ for values of $s$ that were not used as collocation points. These errors can be reduced in magnitude through mesh refinement, utilising higher-order singularities \citep{smithdsl}, or higher-order basis functions \citep{smithvecil}, however they cannot be eliminated entirely at the ends of the body without resorting to replacing the line distribution by a surface distribution, which incurs considerably higher computational expense.
In an earlier computational study utilising singular Stokeslet distributions, \citet{smithdsl} found that cilia could not be arranged in arrays with a spacing of less than $0.3L$, equivalent to approximately $1.8 \smum$, without significant numerical errors occurring in the computational solution---a principal difficulty being that collocation points on the cilia may be close together, particularly during the recovery stroke. This limited the applicability of the technique in determining the effect of cilia density at realistic concentrations.

In this study, we test the RSM without the use of higher-order singularities, and without the use of surface distributions, in order to model a densely-packed patch of cilia. 
The array is modelled as a square lattice of $M=M_1 \times M_2$ cilia, for $1\leqslant l \leqslant M_1$ and $1\leqslant m \leqslant M_2$; at time $t$ the velocity field is given by,
\begin{eqnarray}
\bs{u}(\bs{x},t)&=&\frac{1}{8\pi\mu}\sum_{l=1}^{M_1} \sum_{m=1}^{M_2} \int_0^1 \bs{\mathsf{B}}^\epsilon (\bs{x},\bs{\xi}^{l,m}(s,t))\cdot \bs{f}^{l,m}(s,t) \; \mathrm{d} s \mbox{,} \nonumber      \\
                &=&\frac{1}{8\pi\mu}\sum_{l=1}^{M_1} \sum_{m=1}^{M_2} \sum_{r=1}^N  \bs{f}^{l,m}_r(t) \cdot \int_{(r-1)/N}^{r/N} \bs{\mathsf{B}}^\epsilon (\bs{x},\bs{\xi}^{l,m}(s,t)) \; \mathrm{d} s \mbox{,} \label{ciliaU}
\end{eqnarray}
where the constant vector $\bs{f}^{l,m}_r(t)$ denotes the force on the $r$th segment of the cilium in the $(l,m)$-position in the array, at time $t$.

We use the mathematical formulation of the tracheal cilium beat cycle given by \citet{fulb}, based on the micrograph observations of \citet{sanders}. 
The mathematical specification has the following form:
\begin{equation}
\bs{\xi}(s,t)=\sum_{p=1}^6 \bs{\alpha}_p(s) \cos(pt) + \bs{\beta}_p(s) \sin(pt) \mbox{,} \\
\end{equation}
where the functions $\bs{\alpha}_p(s)$ and $\bs{\beta}_p(s)$ are cubic polynomials, the coefficients being given in \citet{fulb}. 
Rather than impose a metachronal wave on the cilia, we specify that they beat in synchrony, since fully-developed metachronal waves are generally more typical of larger cilia arrays.
Hence the position vector of the cilium in the $(l,m)$-position is simply
\begin{equation}
\bs{\xi}^{l,m}(s,t)=\bs{\xi}(s,t) + (l-1)a\bs{e}_1 + (m-1)a\bs{e}_2\mbox{,}
\end{equation}
where the parameter $a$ is the cilia spacing.

In the numerical implementation, the flow field is calculated at $t_n=n\delta t$, where $n=1,\ldots,N_T$ and $\delta t=2\pi/N_T$. 
We find that results for the volume flow rate converge to three significant figures using $N_T=40$. 
As discussed in \ref{appendixAnalytic} and \ref{testing}(\ref{rods}), we choose the regularisation parameter based on the cilium radius to length ratio, so that 
$\epsilon=0.1/6=0.0167$.
Results are calculated using the University of Birmingham's BlueBEAR Opteron cluster, however the most computationally expensive simulation of 169 cilia requires only 8 hours 20 minutes of CPU time, and so could be performed on a desktop PC.

To monitor the accuracy of the numerical solution for closely-approaching cilia, we use a similar technique to \citet{gillies09} and evaluate the
post-hoc collocation error at each timestep $t=t_n$, denoted $\mathcal{E}_n^{l,m}$. This is given by
\begin{equation}
\mathcal{E}_n^{l,m}=\sqrt{ \int_0^1 | \bs{u}(\bs{\xi}^{l,m}(s,t_n)) - \dot{\bs{\xi}}^{l,m}(s,t_n) |^2 \; \mathrm{d} s  } \mbox{,}
\end{equation}
where the integral is calculated numerically using the points $s_p=(p-1/2)/120$ for $1 \leqslant p \leqslant 120$. For all results presented, we used $N=20$ elements per cilium, and verified that the error did not exceed $\mathcal{E}_n^{l,m}<0.04$ on any cilium $(l,m)$ or any timestep $n$. Example results showing the post-hoc comparison are given in figure~\ref{vCilTest}. The discretised matrix $A$ has similar favourable properties to the matrix resulting from the discretisation using a singular kernel, with the diagonal entries being relatively large. The condition numbers calculated for the multi-cilia results were approximately $1.3 \times 10^{-3}$.

Figure~\ref{vFieldCilia} shows examples of the three-dimensional flow field computed with the model at four points in the beat cycle. The rapid decay of the flow magnitude in both horizontal and vertical directions is evident. The tendency of a patch of cilia to draw fluid `in' from each side to the rear, and push fluid `out' to the side in front is evident, as may be anticipated from the far-field form of the image system \citep{bl71sto}. 

An important function of cilia is the transport of liquid, and it is of interest to determine how the variation in cilia density observed in different biological systems might affect transport. In order to quantify this, we calculated the mean volume flow rate $\bar{Q}$, using the formula described in \ref{vfmCalc}. To examine the different fluid dynamic mechanisms at work, we simulate different cilia densities for flow driven by a finite patch in two ways: (a) keeping the number of cilia constant and varying the spacing through the size of the patch, (b) keeping the size of the patch constant, and varying the spacing through the number of cilia. Results are shown in figure~\ref{vfrFig}.

For a $3\times 3$ cilia patch, as the cilia are brought closer together they hydrodynamically interact, reducing the fluid transport (figure~\ref{vfrFig}a). Because of this interaction effect, for a patch of fixed dimensions, increasing the cilia number from $3\times 3$ to $13\times 13$ results only in a $\simeq 50\%$ increase in fluid transport, and moreover $9\times 9$ cilia give approximately the same transport as a $13\times 13$ array (figure~\ref{vfrFig}b). The common point in figures~\ref{vfrFig}a, b corresponds to cilia spacing $0.5L$, for which the nondimensional mean volume flow rate is approximately $1.45$. These results were computed with a prescribed beat pattern and frequency---in the biological system, these will vary depending on the flow field produced by the other cilia, as discussed in \S\ref{future}.

\subsection{The modelling of a choanoflagellate}\label{choanoSection}
Marine choanoflagellates are the nearest unicellular relatives of the animal kingdom, and exhibit a very diverse array of structures and behaviours that have yet to be fully explained biologically.
Their feeding and swimming behaviour involves the beating of a single flagellum, and has been the subject of previous fluid mechanical modelling utilising slender body theory \citep{ormejfm}. For recent review of the biology, see \citet{leadbeater09}, and for details on flagellar movement and fluid transport, see \citet{pettitt02}.
Certain species exhibit a silica surrounding basket-like structure, known as a \textit{lorica}, made up of a network of silica strips, termed \textit{costae}. The lorica has been hypothesised to perform a number of roles including protecting the cell, reducing the translational velocity of
free-swimming cells, and directing feeding currents to its mouth. Simulating these behaviours is a subject for considerable future study; in this paper we simply demonstrate that with the boundary element regularised Stokeslet method
it is computationally inexpensive to simulate such flows, and we make an initial study of the effect of surrounding the cell with a simple lorica structure.

Both the flagellum and the 16 costae are modelled as filaments, the cell body being modelled as a sphere with $6\times 4 \times 4$ elements, with $N=30$ constant-force elements for the flagellum, and
$N=8$ constant-force elements for each costa. The costae were modelled less precisely than the flagellum, since their role is simply to provide a hydrodynamic drag.
For these simulations we choose $\epsilon = 0.01$, corresponding to a flagellum aspect ratio of $0.1\smum/10\smum$. The same regularisation parameter was also used for the sphere surface, since this has been verified to give accurate results, as shown in \ref{testing}, and for the costae. Simulations for a single timestep were performed on a desktop PC, requiring under ten minutes of CPU time.

\begin{figure}
\begin{center}
\includegraphics{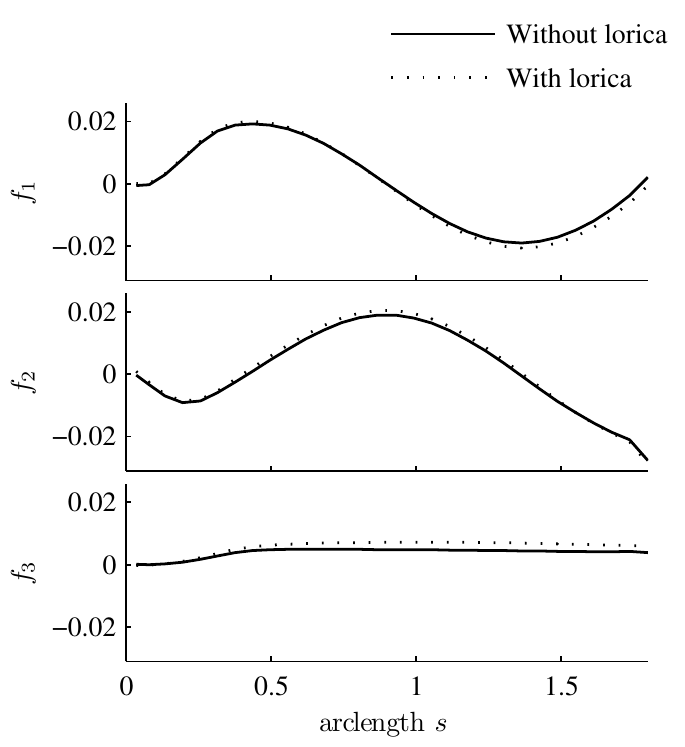}
\end{center}
\caption{Choanoflagellate flagellar force components per unit length versus arclength $s$, with and without the 16 \textit{costae} lorica structure.}\label{choanoForce}
\end{figure}

Our simplified model for the choanoflagellate flagellum, based on that in \citet{higdon79he}, consists of a sphere of radius $\alpha=0.15$, located at $(0, 0, h_0)$, where $h_0=0.4$. The flagellum is described by
\begin{eqnarray}
\xi_1(\hat{s},t) & = & E_{\mathrm{F}}(\hat{s})\cos(\kappa \hat{s} - t)\mbox{,} \nonumber \\
\xi_2(\hat{s},t) & = & E_{\mathrm{F}}(\hat{s})\sin(\kappa \hat{s} - t)\mbox{,} \nonumber \\
\xi_3(\hat{s},t) & = & h_0+\hat{s} \mbox{,} \quad (0<\hat{s}<1) \mbox{,}
\end{eqnarray}
where $t$ is scaled with respect to inverse radian frequency, and $\hat{s}$ is scaled with respect to flagellum extent rather than length. The wavenumber $\kappa=2\pi$, so that the wavelength is unity. The envelope parameter $E_{\mathrm{F}}(\hat{s})=1-\exp(-(\kappa_E\hat{s})^2)$, where $\kappa_E=\kappa$. The coordinate $\hat{s}$ is not an arclength parametrisation, and so it is necessary to make a change of variable to arclength parameter $s$ in order to calculate the force per unit length shown in figure~\ref{choanoForce}.

The silica lorica is constructed using a simplified model, based on the more complex family of geometric models of \citet{leadbeater09}, in which it was shown that there are two sets of costae that run vertically or spiral around the axis of the cell, creating a basket-like structure. We defined a family of $N_C$ vertical strips, given by $\bs{\xi}^{V;m}(\hat{s})$ for $m=1,\ldots N_C$,
\begin{eqnarray}
\xi_1^{V;m}(\hat{s}) &=& E_{\mathrm{C}}(\hat{s})\cos(2\pi m/N_C) \mbox{,} \nonumber \\
\xi_2^{V;m}(\hat{s}) &=& E_{\mathrm{C}}(\hat{s})\sin(2\pi m/N_C) \mbox{,} \nonumber \\
\xi_3^{V;m}(\hat{s}) &=& 2\hat{s} \mbox{,} \quad (0<\hat{s}<1) \mbox{.} \label{lorverteq}
\end{eqnarray}
The family of $N_C$ spiralling strips, denoted $\bs{\xi}^{S;m}(\hat{s})$ for $m=1,\ldots N_C$, was then generated by changing the argument of the trigonometric functions in equation~\eqref{lorverteq} to $2\pi (m/N_C-\hat{s})$. For this study we chose the envelope function $E_C(\hat{s})=3.6\hat{s}(1-\hat{s})$, and a total of $2N_C=16$ costae.

\begin{figure}
$
\begin{array}{ll}
\mbox{(a)} & \mbox{(b)} \\
\includegraphics{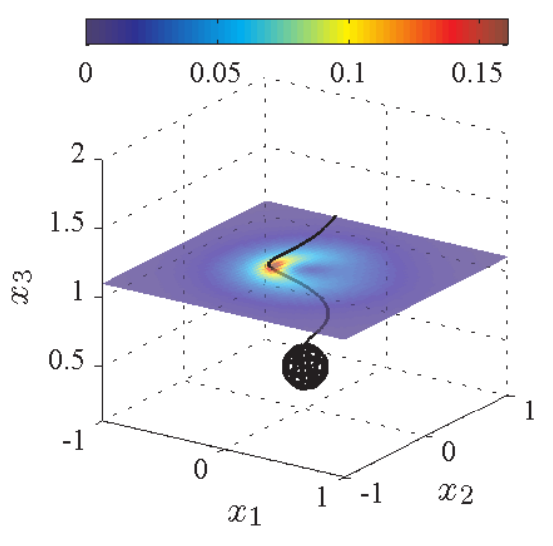}
&
\includegraphics{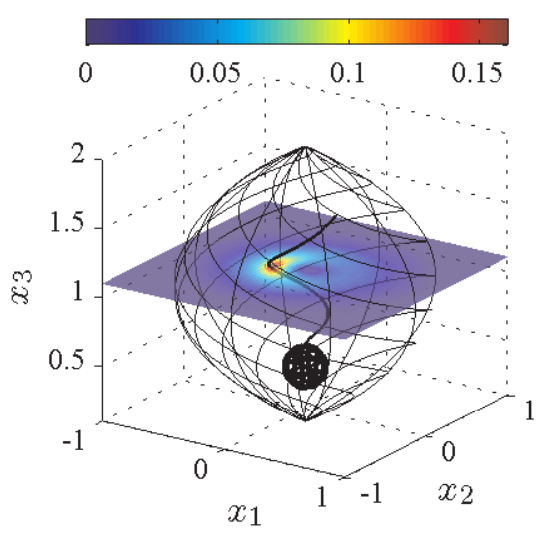} \\
\mbox{(c)} & \mbox{(d)} \\
\includegraphics{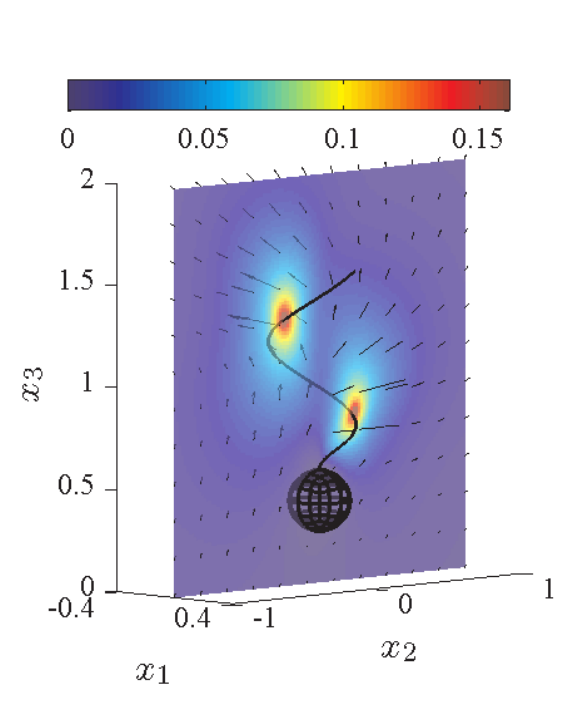}
&
\includegraphics{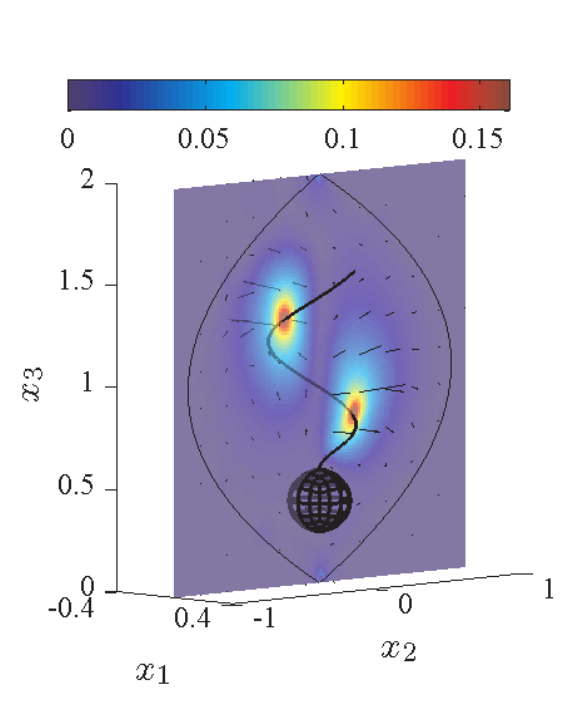}
\end{array}
$
\caption{Choanoflagellate (a,c) without, (b,d) with, a lorica structure consisting of 16 silica {costae}, showing the flow field magnitude (colour surfaces) and direction (arrows in c,d).}\label{choano}
\end{figure}

\begin{figure}
$
\begin{array}{ll}
\mbox{(a)} & \mbox{(b)} \\
\includegraphics{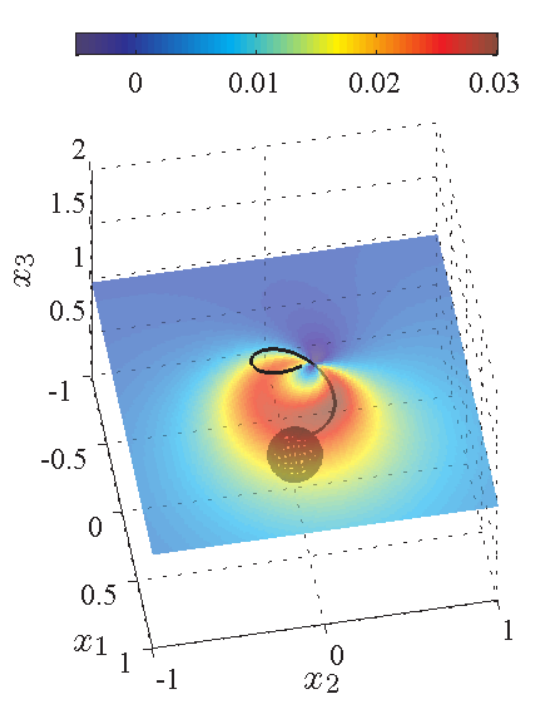}
&
\includegraphics{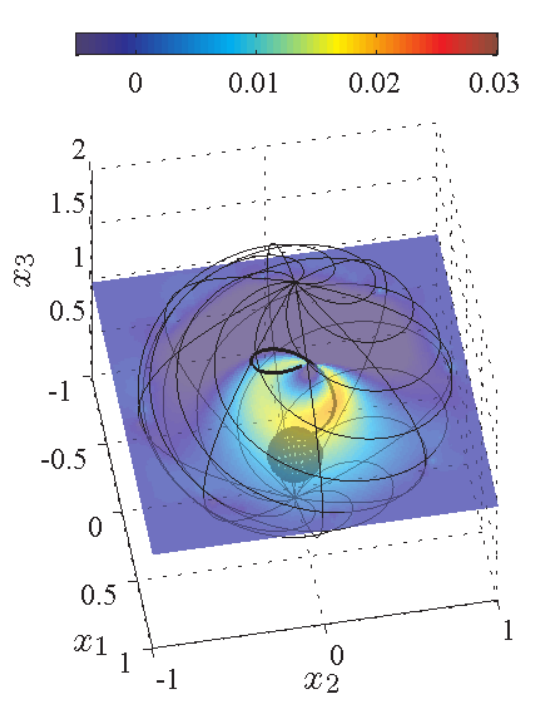}
\end{array}
$
\caption{Choanoflagellate (a) without, (b) with, a lorica structure consisting of 16 silica {costae}, showing the flow field $u_3$ generated.}\label{choanoU3}
\end{figure}

Figure~\ref{choano} shows the instantaneous fluid velocity field, computed on horizontal and vertical surfaces, both with and without the lorica. While the flow close to the flagellum is relatively unchanged by the presence of the lorica, it is apparent from figure~\ref{choano}(c,d) that the remainder of the flow field is strongly suppressed in magnitude. Figure~\ref{choanoU3} shows the vertical component of the velocity field $u_3$ only, on a horizontal cross-section. The presence of the lorica reduces this component in magnitude by nearly 50\% within the lorica interior, and again almost completely abolishes any external flow. By contrast, the hydrodynamic interaction of the flagellum and lorica is relatively weak, with figure~\ref{choanoForce} showing that the force density components are almost unchanged by its presence.

\section{Discussion and future work}\label{future}
A simple modification of the regularised Stokeslet method was presented, based on decoupling the force discretisation and boundary integration. This reduced the number of degrees of freedom required to compute accurate solutions, and removed the need for an empirical parameter relating regularisation parameter and the discretisation mesh for the force.
The implementation was designed to be the most elementary possible, to emphasise its ease of use in biological fluid mechanics problems.
For the Stokes' law test example, a considerable reduction in computational cost was obtained, with approximately 1000-fold fewer kernel evaluations being required to obtain accuracy to within $1\%$, and a 144-fold reduction in the number of degrees of freedom $N$. For problems involving
$M$ bodies, the storage and linear solver costs are at least $O(N^2M^2)$, hence
a reduction in the number of degrees of freedom for many-body problems will be very useful for the modelling of complex biological flows. 

The regularisation parameter $\epsilon$ has two distinct roles in models that employ surface distributions and line distributions. For surface distributions, $\epsilon$ plays the role of a tunable parameter, with smaller values giving more accurate solutions at the expense of more abscissae being required if numerical quadrature is used. It is sufficient to choose $\epsilon$ small enough that the regularisation error is negligible; \citet{cortez05} showed for the blob function used in this paper that this error is $O(\epsilon^2)$, and in this study we found that for the test case of a translating sphere, a value of $\epsilon=0.01$ gave errors of less than $1\%$ with $162$ degrees of freedom. For line distributions, $\epsilon$ is instead equivalent to a physical parameter of the filament---the slenderness ratio, as shown in \ref{appendixAnalytic}. A detailed asymptotic analysis of the slender body theory shall follow this paper.

The technique is capable of providing accurate results for denser arrays of cilia than have previously been possible, 
without recourse to higher order solutions or basis functions. Arrays of up to 169 cilia, with spacing $0.08$ times their length could be simulated for a 40 timestep cycle in approximately 8 hours with a single CPU. It was found that for prescribed cilia movement, cilia density of less than $0.125$ times their length does not produce significant increases in transport, due to hydrodynamic interaction obliterating any gains resulting from there being more cilia---it is likely that the role of realistically-high cilia densities
will only be fully understood through more complex fluid-structure interaction models which take into account more biophysical details of the system in question.

For discussion of the role of cilia in reproduction, see
\citet{faucidillon}. An additional complication when considering airway cilia is the effect of the serous--mucous bilayer, and possible pressure gradients which may exist due to surface tension, osmosis or both. Pressure gradients will have important effects on fluid transport \citep{mitran07,smithrpn}, and cilia penetration of the mucous layer may allow continued transport when many cilia are inactive \citep{fulb}.

The technique was also used to simulate the effect of the complex silica lorica structure of a marine choanoflagellate on the flow field and force density, with minimal computational cost. 
The lorica does not significantly affect the force distribution on the beating flagellum, but does significantly reduce the far-field flow produced. Future models which consider force and torque-balance for a free-swimming cell, in addition to ambient currents, and other features of the cell geometry such as the collar, may provide insight into the role and function of loricae, and the diversity that exists in choanoflagellate species. 

In line distribution representations, the force density depends on the regularisation parameter---in particular it can be estimated as $\sim \log(1/\epsilon)$ as $\epsilon \rightarrow 0$. For free-swimming problems, for example the work of \citet{gillies09}, the swimming speed depends through force and torque balance on the force density, and hence on the value chosen for the regularisation parameter. As shown in \ref{testing}\ref{rods}, this parameter may be chosen based on the slenderness ratio. This observation is relevant only to line distribution models of flagella, and not surface distributions.

The calculations using line distributions were less accurate than those for translation of a sphere, however the method performed with similar accuracy to our previously-published studies for cilia motion, and superior accuracy for dense arrays.
Future work may involve the use of higher-order solutions, higher-order basis functions and non-uniform meshes to reduce the end error for slender bodies. 
Recently, \citet{smithvecil} presented a slender body theory for modelling linear viscoelastic flow, which exploits computational methods developed using singular solutions of the Stokes flow equations. A limitation is that errors in the slender body velocity field are propagated through the solution, which can lead to numerical instability if the Deborah number is larger than approximately $0.06$ for the problem considered. The RSM, possibly combined with higher order Green's functions, may provide a more accurate solution, particularly for problems with multiple cilia, and hence may extend the range of Deborah number that may be modelled.

Other researchers have regularised the slender body theory equations differently: the `Lighthill theorem' \citep{lighthill76}, later exploited by \citet{dresdner} and by \citet{gueron92,gueron99} in the form of the Lighthill-Gueron-Liron theorem, are based on singular Stokeslet distributions, but employ integration to remove the near-field singularity, so that the local part of the integral equation is replaced by resistance coefficients. A related but different approach is evident in the work of \citet{tornberg} on the modelling of complex suspensions of flexible fibres. These techniques have proved very successful in providing an efficient fluid mechanics solver as part of a model of cilia or fibre interactions with the ambient fluid \citep{tornberg,gueron99,gueron01cilia}. An advantage of the regularised Stokeslet method is that the modelling of particles, filaments and surfaces is unified within a single mathematical framework. The present study is intended to expand the scope of the technique to problems which may previously have been computationally difficult, while retaining its ease of implementation. It should be noted that with sophisticated modern parallel computation, direct solution of the fluid flow and solid mechanics equations for large arrays of cilia is possible, see for example \citet{mitran07}.

In this study we considered the flow driven by prescribed cilia and flagellar motions. In order to understand and interpret phenomena such as the effects of viscosity, cilia coordination and cooperation, axoneme ultrastructure, accessory fibres and their effect on beat pattern and frequency, it is necessary to couple the fluid mechanics to models of elasticity and active bending moment generation. Important work has already been done on this by a number of authors, including those discussed above. The framework presented here may assist with the associated fluid mechanics computations, particularly for problems with many bodies, or where there is a complex geometry to be discretised,
for example the surface of the ciliated ventral node of the developing embryo \citep{hirokawa09,smithnodal}, or the interaction of sperm cells with the female reproductive tract.

\section{Acknowledgements}
I thank my colleagues, particularly Professor John Blake, who introduced me to the biological fluid mechanics of cilia and flagella,
and along with Dr Eamonn Gaffney continue to provide invaluable research training, mentoring and guidance.
I also thank Dr Jackson Kirkman-Brown (Centre for Human Reproductive Science, Birmingham Women's Hospital and University of Birmingham) for valuable guidance, and for the opportunity to work with real cilia and flagella first-hand, and Mr Hermes Gad\^{e}lha and Mr Henry Shum (University of Oxford) for fascinating conversations. I also thank Dr Dov Stekel and Dr Barry Leadbeater (University of Birmingham) for discussions on choanoflagellates. This work was funded by the MRC (Special Training Fellowship G0600178).

\appendix{The volume flow rate due to a point-force in the vicinity of a no-slip boundary}\label{vfmCalc}
The pumping effect of a cilium in the $x_1$ direction can be quantified using the volume flow rate $Q$,
\begin{equation}
Q(t)=\int_{x_3=0}^{\infty} \int_{x_2=-\infty}^{\infty} u_1(\bs{x,t}) \; \mathrm{d}x_2 \mathrm{d}x_3 \mbox{,} \label{qDef}
\end{equation}
the choice of $x_1$ being arbitrary due to conservation of mass.
The mean volume flow rate over one beat cycle $(1/T)\int_0^{T} Q(t) \; \mathrm{d} t$ is denoted $\bar{Q}$.

The volume flow rate due to a Stokeslet singularity $B_{1j}$ near to a no-slip surface was given by \cite{liron78}. Our definition of the Stokeslet differs in that we do not include
the leading $1/(8\pi\mu)$ factor, so in our notation we have
\begin{equation}
\int_{x_3=0}^{\infty} \int_{x_2=-\infty}^{\infty} B_{1j}(\bs{x},\bs{\xi}) \; \mathrm{d} x_2 \mathrm{d} x_3   = 
\begin{cases}
              8\xi_3                              
&   \quad \mbox{ if } \quad  j=1   \mbox{,}\\
              0
&   \quad \mbox{ if } \quad  j=2,3   \mbox{,}
\end{cases}
\label{stokesFlux}
\end{equation}
where $x_1$ and $\xi_1$, $\xi_2$ are arbitrary.

By conservation of mass, the volume flow rate does not depend on $x_1$. Taking $x_1$ and hence $r$ to be large,
we note that the regularisation error $|S_{1j}^{\epsilon}-S_{1j}|=O(\epsilon^2/r^3)$,
and the regularisation error for the other images decays at least as rapidly as $O(\epsilon^2/r^4)$. Hence the regularisation error
$|B_{1j}^\epsilon-B_{1j}|=O(\epsilon^2/r^3)$, the double integral of which
tends to zero as $x_1 \rightarrow \infty$, and it cannot make any contribution to the volume flow rate.
This proves that result \eqref{stokesFlux} holds exactly for the regularised solution $B_{1j}^\epsilon$, and so
combining equations~\eqref{qDef}, \eqref{stokesFlux} and \eqref{ciliaU} we have
\begin{equation}
Q(t)  = \frac{1}{\pi \mu} \sum_{l=1}^{M_1} \sum_{m=1}^{M_2} \int_0^1 {f}_1^{l,m}(s,t) \xi_3^{l,m}    (s,t) \; \mathrm{d} s \mbox{.} \label{vfr}
\end{equation}
In \S\ref{biological}\ref{cilia} we report results in nondimensional units, with $Q$ scaled with respect to $\sigma L^3$.

\appendix{Straight line integrals of the regularised Stokeslet and their relation to slender body theory}\label{appendixAnalytic}
For filaments composed of straight line segments, for example the three-link swimmer \citep{purcell77,becker03,tam07}, equation~\eqref{matrixSetup} reduces to the calculation of straight line integrals. Similarly, provided the discretisation is sufficiently fine, curved line integrals can be approximated by a sum of straight line integrals. These integrals can be calculated explicitly, and utilised in a similar manner to the singular Stokeslet-dipole integrals employed by \citet{higdon79}. While this is not required to implement our computational method, in certain situations it is more efficient that using numerical quadrature, for example the results shown in
\ref{testing}(\ref{rods}) and \ref{biological}(\ref{cilia}).
As the radius of curvature of the relative to filament length decreases, it is necessary to subdivide the elements into multiple straight line segments, and the computational savings
are less. For this reason, we use numerical integrals for the choanoflagellate example of \S\ref{choanoForce}.

For an element centred on $\bs{\xi}(s_r)$, having tangent $\bs{\xi}'(s_r)$, the approximating straight line segment is $\bs{\xi}(s)\approx (s-s_r)\bs{\xi}'(s_r)+\bs{\xi}(s_r)$. The regularised Stokeslet integrals are of the form
\begin{equation}
\bs{\mathsf{D}}(\bs{x},s_r)=\int_{s_r-\delta s}^{s_r+\delta s} \bs{\mathsf{S}}^\epsilon(\bs{x},(s-s_r)\bs{\xi}'(s_r)+\bs{\xi}(s_r)) \; \mathrm{d}s.
\end{equation}
We perform the coordinate transformation $x_i^L=(x_j-\xi_j(s_r))\Theta_{ij}$, where $\bs{\mathsf{\Theta}}$ is a rotation matrix. The first row $(\Theta_{11},\Theta_{12},\Theta_{13})$ is given by the tangent $\bs{\xi}'(s_r)$, with the second and third rows completing an orthogonal basis. The integral becomes, in `local' coordinates,
\begin{equation}
\bs{\mathsf{D}}^L(\bs{x},s_r) = \int_{-\delta s}^{\delta s} \bs{\mathsf{S}}^\epsilon(\bs{x}^L,s\mathbf{e}_1) \; \mathrm{d}s \mbox{.}
\end{equation}
The integral in the original coordinate system is given by $D_{ij}(\bs{x},s_r)=\Theta_{ki} D_{kl}^L(\bs{x}^L,s_r) \\ \Theta_{lj}$. The coordinate translation does not affect $\bs{\mathsf{D}}$ since the integrand
$\bs{\mathsf{S}}$ depends on $\bs{x}$ and $\bs{\xi}$ only through their difference.

The definite integrals in the local coordinate system are $D_{ij}^L(\bs{x}^L,s_r)=[I_{ij}^L(\bs{x},s)]_{-\delta s}^{\delta s}$, where $I_{ij}^L(\bs{x}^L,s)$ are the corresponding indefinite integrals. In concise notation $(x,y,z)$ for $(x_1^L,x_2^L,x_3^L)$ these can be written,
\begin{eqnarray}
I_{11}^L& =&-\left(\frac{x-s}{r_\epsilon}\right)\left(\frac{\epsilon^2}{y^2+z^2+\epsilon^2}-1\right) + 2 \log(s-x+r_\epsilon) \nonumber \mbox{,} \\
I_{22}^L& =&-\left(\frac{x-s}{r_\epsilon}\right)\left(\frac{\epsilon^2+y^2}{y^2+z^2+\epsilon^2}\right) + \log(s-x+r_\epsilon) \nonumber \mbox{,} \\
I_{33}^L& =&-\left(\frac{x-s}{r_\epsilon}\right)\left(\frac{\epsilon^2+z^2}{y^2+z^2+\epsilon^2}\right) + \log(s-x+r_\epsilon) \nonumber \mbox{,} \\
I_{12}^L& =& I_{21}^L = \frac{y}{r_\epsilon} \nonumber \mbox{,}  \quad I_{13}^L = I_{31}^L = \frac{z}{r_\epsilon} \nonumber \mbox{,}         \\
I_{23}^L& =& I_{32}^L = - \left(\frac{x-s}{r_\epsilon}\right) \left(\frac{yz}{y^2+z^2+\epsilon^2}\right) \mbox{.}
\end{eqnarray}
with $r_\epsilon^2 = (x-s)^2+y^2+z^2+\epsilon^2$.

The diagonal matrix entries, occurring when $\bs{x}=\bs{\xi}(s_r)$, require higher-order numerical quadrature than the other entries since the kernel is nearly singular at this point, and hence are particularly suited to analytic evaluation. 
These `local' integrals are $\int_{-\delta s}^{\delta s} \bs{\mathsf{S}}^\epsilon(\bs{0},s\mathbf{e}_1) \;\mathrm{d}s$, the components of which do not depend on $\bs{x}$ or $s_r$. They have explicit form
\begin{eqnarray}
D_{11}^{\mathrm{Local}} & = & 4\; \mathrm{arctanh}(1+(\delta s/\epsilon)^2)^{-1/2} \nonumber \mbox{,} \\
D_{22}^{\mathrm{Local}} & = & D_{33}^{\mathrm{Local}} = 2(1+(\delta s/\epsilon)^2)^{-1/2} + 2\; \mathrm{arctanh}(1+(\delta s/\epsilon)^2)^{-1/2}  \nonumber \mbox{,} \\
D_{12}^{\mathrm{Local}} & = & D_{21}^{\mathrm{Local}} = D_{13}^{\mathrm{Local}}=D_{31}^{\mathrm{Local}}=D_{23}^{\mathrm{Local}}=D_{32}^{\mathrm{Local}}=0 \mbox{.}
\end{eqnarray}
The local integrals also provide a link from the numerical implementation to classical slender body theory. 
Taking $\epsilon$ as a proxy for slender body radius, and assuming that the length scale on which the force density varies, and the radius of curvature, are small in comparison with $\epsilon$, then it is reasonable to take $\epsilon \ll \delta s \ll 1$. The `local' integrals then give the dominant contribution of the Stokeslet distribution in the calculation of the force density, corresponding to the resistance coefficients of Gray and Hancock theory.
Moreover, rescaling $\hat{\epsilon}=\epsilon/(2\delta s)$, we have the asymptotic dependence on $\log(1/\hat{\epsilon})$ for $\hat{\epsilon} \ll 1$, as found by \citet{batchelor70}.

\appendix{Test problems}\label{testing}
Two test problems are considered: the translation of a fibre normal to its axis, and the translation of a sphere, both in an infinite fluid. The latter problem was previously considered extensively by 
\cite{cortez05}.

\subsection{Test case 1: force on an axially-translating slender body}\label{rods}
A classical slender body theory result \cite[see for example][noting the slightly different notation]{batchelor70} is that the magnitude of the force on a rod of length $l$ and radius $a$, translating with velocity $u$ parallel to its 
axis is, to leading order $\sim 2\pi \mu u l/\log(l/a)$. Nondimensionalising with force scale $\mu u l$, we calculate a numerical solution using a line distribution of regularised Stokeslets, with the regularisation parameter 
chosen as $\epsilon=a/l$. We compare the two numerical implementations given in equations~\eqref{simpleImp} and \eqref{refinedImp} with $a/l=0.01$, and examine the force at the midpoint of the rod. For the latter case, integrals are calculated analytically, as described in \ref{appendixAnalytic}.
The `boundary element' implementation~\eqref{refinedImp} gives a relative error of less than $3.5\%$ compared with the asymptotic estimate for $N=5,10,20,40,80,160$. The original 
implementation~\eqref{simpleImp} gives relative error of over $35\%$ for $N=10$, and it is necessary to use $N=80$ degrees of freedom to give similar accuracy to the boundary element version.

Benchmarking the two algorithms on a desktop PC with $N=20$ for the boundary element version and $N=80$ for the original implementation, the former requires less than half the CPU time,
using analytic kernel integration. If we repeat the calculation with $M=5$ equally-spaced parallel rods,
the boundary element version requires approximately one sixtieth of the CPU time compared with the original implementation. The reason for the significant saving is that the
matrix setup cost is $O(M^2N^2)$, and the direct solution cost is $O(M^3N^3)$.

\subsection{Test case 2: force on a translating sphere}\label{sphere}
\begin{figure}
\[
\begin{array}{ll}
\mbox{(a) }6\times 4\times 4 \mbox{ elements}& \mbox{(b) }6\times 12\times 12 \mbox{ elements}\\
\includegraphics{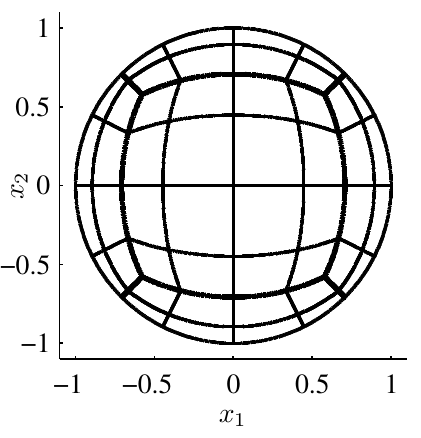} & \includegraphics{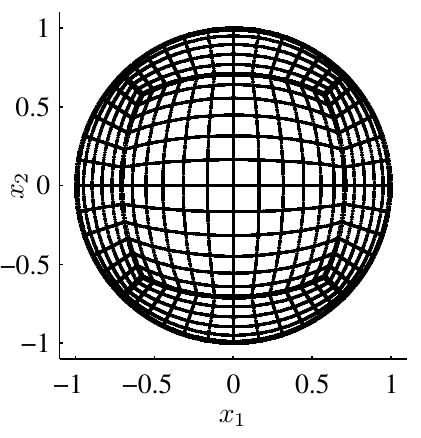}
\end{array}
\]
\caption{Examples of computational meshes used for the test case in \ref{testing}\ref{sphere}, table~\ref{sphereConvergence}, 
with (a) $6\times 4 \times 4$ elements and (b) $6\times 12 \times 12$ elements. The mesh shown in (a) is used in to represent the choanoflagellate cell body in the results shown in 
\S\ref{biological}\ref{choanoSection}, figures~\ref{choano}, \ref{choanoU3} and \ref{choanoForce}.} \label{sphereMesh}
\end{figure}

To test the method with a surface distribution, we compute the drag force $F=\bs{u} \cdot \int_S \bs{f} \; \mathrm{d}S_{\bs{\xi}}$ on a translating sphere in an infinite fluid. 
Stokes' law gives this as exactly
$F=6\pi \mu u a$, where $a$ is the sphere radius. 
Nondimensionalising with force scale $\mu u a$, we compare the boundary element method~\eqref{refinedImp} with the results of 
\cite{cortez05}. We use the same geometric discretisation, projecting the faces of a cube onto its surface, and subdividing these faces with a square mesh, as shown in figure~\ref{sphereMesh}.
The square mesh provides a natural parametrisation of each element, which we denote $(\alpha,\beta)$.
For the boundary element method, numerical integration is performed using
two dimensional Gauss-Legendre quadrature with the parametrised coordinates, taking $12\times 12$ points for the `near-singular' cases $q=r$ in equation~\eqref{refinedImp}
and $4 \times 4$ points otherwise.
The surface metric is calculated as the magnitude of the vector product $|\partial \bs{\xi}/\partial \alpha \wedge \partial \bs{\xi} / \partial \beta|$, as described in \citet{bemlib}.
All of the collocation and quadrature points lie on the 
curved surface of the sphere.

For problems involving surface integrals, refining the mesh by halving the element width results in four times the number of degrees of freedom,
16 times the matrix setup and storage cost, and at least the same cost increase for an iterative linear solver. With a conventional direct solver, the cost increases with $O(N^3)$, i.e.\ 64 times.
For this reason, a method that gives accurate results with a relatively coarse mesh is beneficial.
Consistent with \citet{cortez05}, we compute solutions using the generalised minimum residual method with zero initial guess, although
for the cases $N \leqslant 6\times 12 \times 12$; solution by $LU$ factorisation is possible in a few seconds on a desktop PC.

Table~\ref{sphereConvergence}a is based on the results with regularisation parameter $\epsilon=0.01$
reported by \cite{cortez05} using equation~\eqref{simpleImp}, although we include relative percentage error and the number of scalar kernel evaluations necessary to construct the matrix $A$. In order to obtain
a relative error of less than $1\%$ it is necessary to use a mesh with $6\times 36\times 36$ patches, giving $23328$ scalar degrees of
freedom for the unknown force components, the matrix setup requiring $5.44\times 10^{8}$ kernel evaluations.

Table~\ref{sphereConvergence}b
shows results calculated using equation~\eqref{refinedImp}, with coarser meshes and fewer degrees of freedom. To obtain similar accuracy, a $6\times 3 \times 3$
mesh with 162 DOF is sufficient. While the matrix setup cost per DOF is greater due to the higher-order
quadrature, the greatly reduced DOF results in the number of kernel evaluations being three orders of
magnitude smaller. The matrix solution cost is reduced similarly.

The effect of varying the regularisation parameter $\epsilon$ on the predicted drag was also examined. Using $N=6\times 4\times 4$ and constant force discretisation, the relative error for $\epsilon=0.05$ was $1.4\%$, reducing to $0.63\%$ for $\epsilon=0.01$. The error remains at approximately $0.62\%$ for $\epsilon=0.005$ and $0.0025$, implying that this is a satisfactory value---the remaining error arises from the force discretisation. \citet{cortez05} showed that the error associated with the regularisation parameter can be estimated as $O(\epsilon^2)$. Hence a value of $\epsilon$ that gives accurate results for the sphere problem will in general give accurate results for any solid surface for which the length scales are significantly larger than $\epsilon$. Moreover, once $\epsilon$ has been chosen sufficiently small, accurate results may be computed for any values of $\epsilon$ that are smaller, provided that the kernel quadrature is computed accurately.

\begin{table}
(a), Original method \vspace{0.1in}\\
\(
\begin{array}{lllll}	\hline
N                     & \mbox{DOF}     & \mbox{matrix entries}   & \mbox{no. kernel}       & \mbox{rel. \% error}  \\
                      &                   &                         & \mbox{evaluations}      & \mbox{in drag}       \\ \hline
6\times 12 \times 12  & 2592              &  6.72\times 10^{6}      & 6.72\times 10^{6}       &   12.6    \\
6\times 24 \times 24  & 10368             &  1.07\times 10^{8}      & 1.07\times 10^{8}       &   2.76    \\
6\times 36 \times 36  & 23328             &  5.44\times 10^{8}      & 5.44\times 10^{8}       &   0.849    \\
6\times 48 \times 48  & 41472             &  1.72\times 10^{9}      & 1.72\times 10^{9}       &   0.265    
\end{array}
\) \\ \vspace{0.2in}\\
(b), Boundary element method \vspace{0.1in}\\
\(
\begin{array}{lllll}\hline
N                     & \mbox{DOF}     & \mbox{matrix entries}   & \mbox{no. kernel}       & \mbox{rel. \% error}  \\
                      &                   &                         & \mbox{evaluations}      & \mbox{in drag}       \\ \hline
6\times 3 \times 3              & 162         &  26244			&   482112                      &    0.827   \\
6\times 4 \times 4              & 288         &  82944                  &   1.44\times 10^{6}   &    0.626   \\
6\times 6 \times 6              & 648         & 419904                  &   6.97\times 10^{6}   &    0.431   \\
6\times 9 \times 9              & 1458        & 2.12\times 10^{6}       &   3.48\times 10^{7}   &    0.320   \\
6\times 12 \times 12            & 2592        & 6.72\times 10^{6}       &   1.08\times 10^{8}   &    0.279   
\end{array}
\)
\caption{Convergence of the original RSM results given in \cite{cortez05} and the boundary element implementation equation~\eqref{refinedImp}, for the
canonical test problem of the drag force on a translating unit sphere. DOF is degrees of freedom. The regularisation parameter was fixed at $\epsilon=0.01$ in both cases. The
boundary element implementation used $12\times 12$--point Gauss-Legendre quadrature for `near-singular' integrals, corresponding to $q=r$ in equation~\eqref{refinedImp},
and $4\times 4$-point quadrature for all other integrals.}\label{sphereConvergence} 
\end{table}


\begin{thebibliography}{}

\bibitem[Ainley et~al., 2008]{ainley08}
Ainley, J., Durkin, S., Embid, R., Boindala, P., and Cortez, R. (2008).
\newblock The method of images for regularised {S}tokeslets.
\newblock {\em J. Comput. Phys.}, 227:4600--4616.

\bibitem[Batchelor, 1970]{batchelor70}
Batchelor, G.~K. (1970).
\newblock Slender-body theory for particles of arbitrary cross-section in
  {S}tokes flow.
\newblock {\em J. Fluid Mech.}, 44:419--440.

\bibitem[Becker et~al., 2003]{becker03}
Becker, L.~E., Koehler, S.~A., and Stone, H.~A. (2003).
\newblock On self-propulsion of micro-machines at low reynolds number:
  Purcell's three-link swimmer.
\newblock {\em J. Fluid Mech.}, 490:15--35.

\bibitem[Blake, 1971]{bl71sto}
Blake, J.~R. (1971).
\newblock A note on the image system for a stokeslet in a no slip boundary.
\newblock {\em Proc. Camb. Phil. Soc.}, 70:303--310.

\bibitem[Blake, 1972]{bl72}
Blake, J.~R. (1972).
\newblock A model for the micro-structure in ciliated organisms.
\newblock {\em J. Fluid Mech.}, 55:1--23.

\bibitem[Burgers, 1938]{burgers38}
Burgers, J.~M. (1938).
\newblock On the motion of small particles of elongated form suspended in a
  viscous liquid.
\newblock {\em Kon. Ned. Akad. Wet. Verhand. (Eerste Sectie)}, 16:113.

\bibitem[Chwang and Wu, 1975]{chwang}
Chwang, A.~T. and Wu, T.~Y. (1975).
\newblock Hydrodynamics of the low-{R}eynolds number flows. {P}art 2. the
  singularity method for {S}tokes flows.
\newblock {\em J. Fluid Mech.}, 67:787--815.

\bibitem[Cisneros et~al., 2007]{cisneros07}
Cisneros, L.~H., Cortez, R., Dombrowski, C., Goldstein, R.~E., and Kessler,
  J.~O. (2007).
\newblock Fluid dynamics of self-propelled microorganisms, from individuals to
  concentrated populations.
\newblock {\em Exp. Fluids}, 43:737--753.

\bibitem[Cisneros et~al., 2008]{cisneros08}
Cisneros, L.~H., Kessler, J.~O., Ortiz, R., Cortez, R., and Bees, M. (2008).
\newblock Unexpected bipolar flagellar arrangements and long-range flows driven
  by bacteria near solid boundaries.
\newblock {\em Phys. Rev. Lett.}, 101:168102.

\bibitem[Clarke et~al., 2006]{clarke}
Clarke, R.~J., Jensen, O.~E., Billingham, J., and Williams, P.~M. (2006).
\newblock Three-dimensional flow due to a microcantilever oscillating near a
  wall: an unsteady slender-body analysis.
\newblock {\em Proc. R. Soc. A}, 462:913--933.

\bibitem[Cortez, 2001]{cortez01}
Cortez, R. (2001).
\newblock The method of regularized stokeslets.
\newblock {\em SIAM J. Sci. Comput.}, 23(4):1204--1225.

\bibitem[Cortez et~al., 2005]{cortez05}
Cortez, R., Fauci, L., and Medovikov, A. (2005).
\newblock The method of regularized stokeslets in three dimensions: Analysis,
  validation and application to helical swimming.
\newblock {\em Phys. Fluids}, 17(031504):1--14.

\bibitem[Dresdner et~al., 1980]{dresdner}
Dresdner, R.~D., Katz, D.~F., and Berger, S.~A. (1980).
\newblock The propulsion by large amplitude waves of uniflagellar
  micro-organisms of finite length.
\newblock {\em J. Fluid Mech.}, 97:591--621.

\bibitem[Fauci and Dillon, 2006]{faucidillon}
Fauci, L. and Dillon, R. (2006).
\newblock Biofluidmechanics of reproduction.
\newblock {\em Ann. Rev. Fluid Mech.}, 38(1):371--394.

\bibitem[Flores et~al., 2005]{flores05}
Flores, H., Lobaton, E., M\'{e}ndez-Diez, S., Tlupova, S., and Cortez, R.
  (2005).
\newblock A study of bacterial flagellar bundling.
\newblock {\em Bull. Math. Biol.}, 67:137--168.

\bibitem[Fulford and Blake, 1986]{fulb}
Fulford, G.~R. and Blake, J.~R. (1986).
\newblock Muco-ciliary transport in the lung.
\newblock {\em J. Theor. Biol.}, 121:381--402.

\bibitem[Gillies et~al., 2009]{gillies09}
Gillies, E.~A., Cannon, R.~M., Green, R.~B., and Pacey, A.~A. (2009).
\newblock Hydrodynamic propulsion of human sperm.
\newblock {\em J. Fluid Mech.}, 625:444--473.

\bibitem[Gray and Hancock, 1955]{grayhan}
Gray, J. and Hancock, G.~J. (1955).
\newblock The propulsion of sea-urchin spermatozoa.
\newblock {\em J. Exp. Biol.}, 32(4):802--814.

\bibitem[Gueron and Levit-Gurevich, 1999]{gueron99}
Gueron, S. and Levit-Gurevich, K. (1999).
\newblock Energetic considerations of ciliary beating and the advantage of
  metachronal coordination.
\newblock {\em Proc. Natl. Acad. Sci. USA}, 96(22):12240--12245.

\bibitem[Gueron and Levit-Gurevich, 2001]{gueron01cilia}
Gueron, S. and Levit-Gurevich, K. (2001).
\newblock A three-dimensional model for ciliary motion based on the internal
  9+2 structure.
\newblock {\em {Proc. R. Soc. Lond. B}}, 268:599--607.

\bibitem[Gueron and Liron, 1992]{gueron92}
Gueron, S. and Liron, N. (1992).
\newblock Ciliary motion modeling, and dynamic multicilia interactions.
\newblock {\em Biophys. J.}, 63:1045--1058.

\bibitem[Hancock, 1953]{hancock53}
Hancock, G.~J. (1953).
\newblock The self-propulsion of microscopic organisms through liquids.
\newblock {\em Proc. R. Soc. B.}, 217:96--121.

\bibitem[Higdon, 1979a]{higdon79}
Higdon, J. J.~L. (1979a).
\newblock A hydrodynamic analysis of flagellar propulsion.
\newblock {\em J. Fluid Mech.}, 90:685--711.

\bibitem[Higdon, 1979b]{higdon79he}
Higdon, J. J.~L. (1979b).
\newblock The hydrodynamics of flagellar propulsion: helical waves.
\newblock {\em J. Fluid Mech.}, 94(2):331--351.

\bibitem[Hirokawa et~al., 2009]{hirokawa09}
Hirokawa, N., Okada, Y., and Tanaka, Y. (2009).
\newblock {Fluid Dynamic Mechanism Responsible for Breaking the Left-Right
  Symmetry of the Human Body: The Nodal Flow}.
\newblock {\em Ann. Rev. Fluid Mech.}, 41:53--72.

\bibitem[Johnson, 1980]{johnson}
Johnson, R.~E. (1980).
\newblock An improved slender-body theory for {S}tokes flow.
\newblock {\em J. Fluid Mech.}, 99(2):411--431.

\bibitem[Leadbeater et~al., 2009]{leadbeater09}
Leadbeater, B., Yu, Q., Kent, J., and Stekel, D. (2009).
\newblock {Three-dimensional images of choanoflagellate loricae}.
\newblock {\em Proc. R. Soc. Lond. B}, 276(1654):3.

\bibitem[Lighthill, 1976]{lighthill76}
Lighthill, M.~J. (1976).
\newblock Flagellar hydrodynamics. {T}he {J}ohn von {N}eumann lecture.
\newblock {\em SIAM Review}, 18(2):161--230.

\bibitem[Liron, 1978]{liron78}
Liron, N. (1978).
\newblock Fluid transport by cilia between parallel plates.
\newblock {\em J. Fluid Mech.}, 86(4):705--726.

\bibitem[Liron and Mochon, 1976]{liron76}
Liron, N. and Mochon, S. (1976).
\newblock The discrete-cilia approach to propulsion of ciliated
  micro-organisms.
\newblock {\em J. Fluid Mech.}, 75:593--607.

\bibitem[Mitran, 2007]{mitran07}
Mitran, S. (2007).
\newblock Metachronal wave formation in a model of pulmonary cilia.
\newblock {\em Comput. Struct.}, 85(11--14):763--774.

\bibitem[Orme et~al., 2003]{ormejfm}
Orme, B. A.~A., Blake, J.~R., and Otto, S.~R. (2003).
\newblock Modelling the motion of particles around choanoflagellates.
\newblock {\em J. Fluid Mech.}, 475:333--355.

\bibitem[Pettitt et~al., 2002]{pettitt02}
Pettitt, M.~E., Orme, B. A.~A., Blake, J.~R., and Leadbeater, B. S.~C. (2002).
\newblock The hydrodynamics of filter feeding in choanoflagellates.
\newblock {\em Europ. J. Protistol.}, 38:313--332.

\bibitem[Phan-Thien et~al., 1987]{phan}
Phan-Thien, N., Tran-Cong, T., and Ramia, M. (1987).
\newblock A boundary-element analysis of flagellar propulsion.
\newblock {\em J. Fluid Mech.}, 185:533--549.

\bibitem[Pozrikidis, 1992]{pozrikidis}
Pozrikidis, C. (1992).
\newblock {\em Boundary integral and singularity methods for linearized viscous
  flow}.
\newblock Cambridge University, Cambridge.

\bibitem[Pozrikidis, 2002]{bemlib}
Pozrikidis, C. (2002).
\newblock {\em A practical guide to boundary-element methods with the software
  library {BEMLIB}}.
\newblock Chapman and Hall/CRC.

\bibitem[Purcell, 1977]{purcell77}
Purcell, E.~M. (1977).
\newblock Life at low {R}eynolds number.
\newblock {\em Am. J. Phys.}, 45(1):3--11.

\bibitem[Ramia et~al., 1993]{ramia}
Ramia, M., Tullock, D.~L., and Phan-Thien, N. (1993).
\newblock The role of hydrodynamic interaction in the locomotion of
  microorganisms.
\newblock {\em Biophys. J.}, 65:755--778.

\bibitem[Sanderson and Sleigh, 1981]{sanders}
Sanderson, M.~J. and Sleigh, M.~A. (1981).
\newblock Ciliary activity of cultured rabbit tracheal epithelium: beat pattern
  and metachrony.
\newblock {\em J. Cell Sci.}, 47:331--341.

\bibitem[Sleigh et~al., 1988]{sleigh88}
Sleigh, M.~A., Blake, J.~R., and Liron, N. (1988).
\newblock The propulsion of mucus by cilia.
\newblock {\em Am. Rev. Respir. Dis.}, 137:726--741.

\bibitem[Smith et~al., 2008a]{smithnodal}
Smith, D.~J., Blake, J.~R., and Gaffney, E.~A. (2008a).
\newblock Fluid mechanics of nodal flow due to embryonic primary cilia.
\newblock {\em J. R. Soc. Interface}, 5:567--573.

\bibitem[Smith et~al., 2007]{smithdsl}
Smith, D.~J., Gaffney, E.~A., and Blake, J.~R. (2007).
\newblock Discrete cilia modelling with singularity distributions: application
  to the embryonic node and the airway surface liquid.
\newblock {\em Bull. Math. Biol.}, 69(5):1477--1510.

\bibitem[Smith et~al., 2008b]{smithrpn}
Smith, D.~J., Gaffney, E.~A., and Blake, J.~R. (2008b).
\newblock Modelling mucociliary clearance.
\newblock {\em Respir. Physiolo. Neurobiol.}, 163:178--188.

\bibitem[Smith et~al., 2009a]{smithvecil}
Smith, D.~J., Gaffney, E.~A., and Blake, J.~R. (2009a).
\newblock Mathematical modelling of cilia-driven transport of biological
  fluids.
\newblock {\em Proc. R. Soc. Lond. A}, 465(2108):2417--2439.

\bibitem[Smith et~al., 2009b]{smithjfm}
Smith, D.~J., Gaffney, E.~A., Blake, J.~R., and Kirkman-Brown, J.~C. (2009b).
\newblock Human sperm accumulation near surfaces: a simulation study.
\newblock {\em J. Fluid Mech.}, 621:289--320.

\bibitem[Tam and Hosoi, 2007]{tam07}
Tam, D. and Hosoi, A. (2007).
\newblock Optimal stroke patterns for {P}urcell{'}s three-link swimmer.
\newblock {\em Phys. Rev. Lett.}, 98:068105.

\bibitem[Tornberg and Shelley, 2004]{tornberg}
Tornberg, A.-K. and Shelley, M.~J. (2004).
\newblock Simulating the dynamics and interactions of flexible fibers in
  {S}tokes flows.
\newblock {\em J. Comp. Phys.}, 196:8--40.

\bibitem[Tuck, 1964]{tuck64}
Tuck, E.~O. (1964).
\newblock Some methods for flows past slender bodies.
\newblock {\em J. Fluid Mech.}, 18:619--635.

\end{thebibliography}
\end{document}